\documentclass[showpacs,aps,prb,twocolumn,superscriptaddress]{revtex4-2}

\usepackage[english]{babel}
\usepackage{graphicx}
\usepackage{hyperref}
\usepackage{epsfig}
\usepackage{color}
\usepackage{xcolor}
\usepackage{psfrag}
\usepackage{float}
\usepackage{tikz}
\usepackage{amsmath}
\usepackage{amssymb}
\newcommand{\bsigma}{\overline{\sigma}}

\DeclareUnicodeCharacter{2009}{\,}
\DeclareUnicodeCharacter{2212}{-}

\begin{document}
%

\title{Stochastic Dynamics of Resonance Electronic Energy Transfer in Bi-Dimensional Overexcited Molecular Ensembles}
%

%
\author{R. Avriller}
\affiliation{Univ. Bordeaux, CNRS, LOMA, UMR 5798, F-33405 Talence, France}
\author{A. March\'e}
\affiliation{Univ. Bordeaux, CNRS, LOMA, UMR 5798, F-33405 Talence, France}
\author{G. Jonusauskas}
\affiliation{Univ. Bordeaux, CNRS, LOMA, UMR 5798, F-33405 Talence, France}
\date{\today}
%

%
\begin{abstract}
We investigate theoretically the stochastic dynamics of Resonance Electronic Energy Transfer (RET), in a bi-dimensional overexcited ensemble of donor and acceptor molecules.
We find that, after initial optical excitation of all the donors, the reaction kinetics is well-described by a non-linear mean-field theory. 
The latter provides a solid way to define and compute an effective rate of RET, even for disordered samples. 
We predict that this effective rate scales as $\left\langle R \right\rangle^{\alpha}$ with $\left\langle R \right\rangle$ the average distance between individual excited donors and their nearest-neighbor acceptor molecules, and $\alpha \in \left\lbrack -6,-2 \right\rbrack$ an exponent depending on the spatial distribution of molecular pairs in the sample.
Using a kinetic Monte-Carlo approach, we show departures from this macroscopic mean-field description arising from fluctuations and spatial correlations between several molecules involved in the RET process.
We expect this prediction to be relevant for both molecular science and biology, where the control and optimization of the RET dynamics is a key issue.
\end{abstract}
\maketitle 
%

%
\section{Introduction}
\label{Intro}
%
%
%
The investigation of how energy dissipates and migrates from one place to another in complex molecular systems, is of paradigmatic importance, in particular for the understanding of various physical, chemical and biological processes \cite{CHIRIOLEBRUN1998320,PhysRevE.82.041905}.
In this context, a key role is played by resonance electronic energy transfer (RET) processes, through which an excitation (labelled $^*$) initially stored onto a donor molecule (D), can be transferred to an acceptor molecule (A), resulting into the RET reaction $D^* + A \rightarrow D + A^*$. 
The corresponding reaction kinetics is usually described by F\"orster resonance energy transfer (FRET) theory \cite{forster1960transfer,10.1117/1.JBO.17.1.011003}, and involves dipole-dipole interactions between two donnor and acceptor molecules. 
The corresponding rate of FRET is obtained from quantum mechanics using Fermi Golden Rule, and is given by $k_{\rm{D-A}}=|V_{\rm{D-A}}|^2 J_{\rm{D-A}} / 2\pi\hbar^2$ \cite{PhysRevLett.92.218301}, with $V_{\rm{D-A}}$ the matrix-element corresponding to the dipole-dipole interaction hamiltonian, $J_{\rm{D-A}}$ the overlap integral between the donor emission spectrum and acceptor absorption spectrum, and $\hbar$ the Planck constant.
This expression can be rewritten in the standard form $k_{\rm{D-A}}=\Gamma_D \left( R_0 / R_{\rm{D-A}} \right)^6$ \cite{forster1960transfer,doi:10.1021/jp900708f}, with $\Gamma_D$ the rate of spontaneous photon emission of the donors, $R_{\rm{D-A}}$ the distance between the D-A molecules, and $R_0$ the F\"orster radius that typically falls in the range of $1-10 \mbox{ nm}$.
%

%
%
%
Despite its long history, the mechanism of FRET is still the object of intense research activities and fruitful debates.  
For instance, some recent theoretical investigations in molecules made of
multiple chromophores, emphasized the central role of non-equilibrium effects and molecular vibrations 
in FRET \cite{doi:10.1021/jp900708f,PhysRevLett.92.218301,doi:10.1063/1.3140273,doi:10.1021/jp0569281,doi:10.1063/1.2977974}.
The latter were shown to be responsible for a vibration-assisted long-range energy-transfer mechanism beyond standard F\"orster theory.
Other studies focused on deriving $k_{\rm{D-A}}$ rigorously, using a non-relativistic quantum-electrodynamics framework to describe the interaction between two molecules, mediated by coupling to virtual states of the electromagnetic reservoir \cite{doi:10.1063/1.1742697,doi:10.1063/1.1579677,ANDREWS1989195,CRAIG198937,CRAIG1992229,doi:10.1063/5.0042684}.
They predicted the existence of a crossover from the standard F\"orster static-regime at short-distances scaling with $k_{\rm{D-A}}\propto R_{\rm{D-A}}^{-6}$, to a long-range retarded radiation-regime mediated by the emission and propagation of real photons scaling with $k_{\rm{D-A}}\propto R_{\rm{D-A}}^{-2}$.
In parallel to these fundamental approaches, a whole range of activities focused on dealing with the question of what is the effective range of RET processes, as well as investigating the role of disorder in the related exciton-transport mechanism \cite{doi:10.1063/1.4861695,doi:10.1146/annurev.physchem.54.011002.103746,doi:10.1021/cr800268n,doi:10.1021/jp4124502}, using for this purpose numerical kinetic Monte-Carlo methods \cite{doi:10.1021/jp0031146,10.1371/journal.pone.0019791,FRAZIER20072422,doi:10.1021/jp202152m,doi:10.1021/jp014549b,doi:10.1021/acs.jpcb.8b07719}.
Most of those studies were motivated by the still open issue of understanding the FRET processes in biological complexes or molecular aggregates, that is supposed to play a key-role in the mechanism of photosynthesis \cite{doi:10.1021/jp003571m,doi:10.1021/jp070111l,doi:10.1021/jp983477u,geddes2006reviews,PhysRevLett.108.218302}.
%
%
%
A revival of those issues occurred recently in a different context, for which it was shown that RET processes could be modified for molecules deposited close to a mirror \cite{doi:10.1063/1.4998459}, or by embedding the donor and acceptor molecules inside electromagnetic microcavities, leading to a new kind of long-range resonance energy transfer mechanism, mediated by vacuum quantum fluctuations of one electromagnetic cavity mode \cite{https://doi.org/10.1002/anie.201703539,PhysRevB.97.241407,https://doi.org/10.1002/anie.202105442,C8SC00171E,reitz2018energy}.
%

%
%
%
The questions of \textit{understanding collective and non-equilibrium effects in RET processes occurring in molecular ensembles or aggregates} (beyond the standard description in terms of single donor and acceptor molecules), as well as the one of defining and computing quantitatively an effective rate of RET from a microscopic model, are thus still lively and open important questions in molecular sciences.
%
%

%
%
%
In this paper, we investigate the stochastic dynamics and collective effects developing in RET reactions. 
We consider the case of a bi-dimensional overexcited molecular system which is prepared in an initial out-of-equilibrium state in which \textit{all donor molecules are brought into their excited-state}.
In Sec.\ref{Sec1}, we introduce a microscopic model of molecular ensemble made of equal concentration of donor and acceptor molecules positioned at the vertex of an arbitrary bi-dimensional network.
We describe the system relaxation, and stochastic dynamics of its  microstates in terms of a chemical master-equation.
The latter is solved numerically exactly using a kinetic Monte-Carlo algorithm.
In Sec.\ref{Sec2}, we derive \textit{exact kinetic equations describing the coarsed-grained dynamics of the average populations of donor and acceptor molecules}. 
In Sec.\ref{Mean-field}, we propose a non-linear mean-field approximation that matches the expected macroscopic limit, and provides a solid approach to define and compute explicitly the rate of RET in any disordered molecular ensemble. 
%
%
We also investigate the role of fluctuations developing out of equilibrium \cite{Fluctuations_Nonequilibrium_Systems,nitzan1974comment,van1992stochastic}, and resulting in deviations from this mean-field approximation. 
Finally, in Sec.\ref{FullKin}, we compute and analyze the complete time-dependent relaxation dynamics of excited donors and acceptors, both from the numerically exact kinetic Monte-Carlo algorithm and macroscopic mean-field approximation.
We show a good agreement of the mean-field approximation with exact results and quantify the role of fluctuations for an accurate description of RET in mesoscopic samples.
%

%
\section{Modeling the stochastic dynamics of RET}
\label{Sec1}
%
%
\subsection{System description}
\label{System}
\begin{figure}[tbh]
\includegraphics[width=1.0\linewidth, angle=-0]{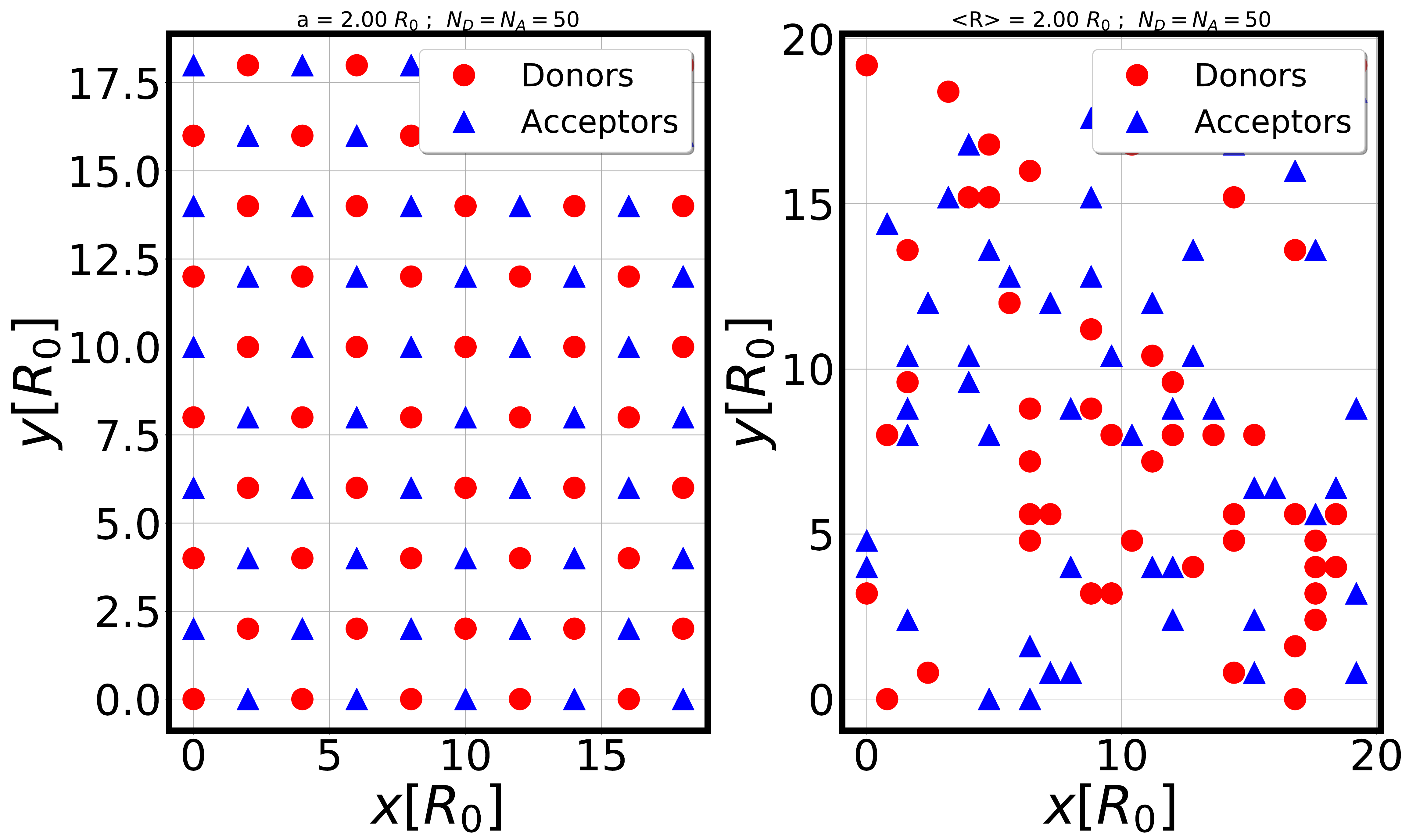}
\caption{
\textcolor{black}{Left: Square lattice with  $a=2.0 R_0$ the distance between two nearest-neighbour molecules.}
Donnor (D) and acceptor (A) molecules are shown as red dots and blue triangles respectively. 
The system contains $N_D=N_A=50$ molecules of type D and A.
\textcolor{black}{Right: Disordered lattice obtained by putting each molecule randomly at the vertex of a regular square lattice of lattice parameter $R_c$.} 
%
%
\textcolor{black}{Parameters are $\left\langle R \right\rangle=2.0R_0$ and $R_c=0.8R_0$.}
}
\label{fig:Fig1}
\end{figure}
%
%
In this section, we introduce a microscopic model of molecular system, made of $N_D$ donor molecules and $N_A$ acceptor molecules. 
%
%
We consider the case of an equimolar solution of molecules, characterized by $N_D=N_A\equiv N/2$.
The molecules are dispersed in a two-dimensional solid-phase.
The molecular positions are labeled by $\vec{r}_{\alpha}$ for 
the $D_\alpha$ molecules with $\alpha \in \left\lbrack |1,N_D| \right\rbrack$, and $\vec{r}_{j}$ for 
the $A_j$ molecules with $j \in \left\lbrack |1,N_A| \right\rbrack$.
%
%
\textcolor{black}{In the following, we consider two cases.
%
%
The first one corresponds to the case of molecules located at the vertex of a regular square lattice (see Fig.~\ref{fig:Fig1}-Left), with $a$ the distance between two nearest-neighbour molecules (and $2a$ the lattice parameter).
The second case corresponds to molecules randomly distributed and forming a random (disordered) lattice (see Fig.~\ref{fig:Fig1}-Right).
This lattice, is made by generating a regular square lattice of lattice parameter $R_c$, a cutoff length corresponding to the minimum distance between two molecules (given by the short-range part of the intra-molecular interaction potential).
Each molecule is then put randomly at one vertex of the square lattice, thus realizing a random lattice in which the average distance between an individual excited-donor and its nearest-neighbor acceptor is $\left\langle R \right\rangle \approx \sqrt{S/ \pi N}$ (proportional to the inverse square root of the chosen molecular concentration). 
}
%
%
%
%
%
%
%
%
%

%
%
In this ensemble, the donor (acceptor) molecules are supposed to be either in their electronic 
ground-state labeled $D(A)$ or in their excited-state written $D^*(A^*)$.
We suppose that initially, an external ultra-fast pump-laser signal is applied that brings all the donors to excited state, leaving the acceptors in their ground-state, so that the initial population for the molecules in the sample is $N_{D^*}(0)=N_D$ and $N_{A^*}(0)=0$.
We aim to describe \textit{how this initial overexcited state that is strongly out-of-equilibrium will relax, either by fluorescence (radiative relaxation), or by RET reactions (non-radiative relaxation)}.
The possible relaxation pathways in this model are described by the following elementary processes 
\begin{eqnarray}
D^*_{\alpha} \rightarrow D_{\alpha} + h\nu
\label{Reac1} \, , \\
A^*_{j} \rightarrow A_{j} + h\nu
\label{Reac2} \, , \\
D^*_{\alpha} + A_{j} \rightarrow D_{\alpha} + A^*_{j}
\label{Reac3} \, ,
\end{eqnarray}
with Eq.\ref{Reac1} describing the relaxation of the $D^*_{\alpha}$ molecule by emission of a spontaneous photon $h\nu$ with rate $\Gamma_D$, and Eq.\ref{Reac2} the similar process for the $A^*_{j}$ molecule, with rate $\Gamma_A$.
The possibility of transferring energy from a $D^*_{\alpha}$ molecule to a $A_{j}$ molecule, is described by the RET process in Eq.\ref{Reac3}, with rate $k_{\alpha j}$.
%
%
The latter rate $k_{\alpha j}\equiv k\left(r=||\vec{r}_\alpha-\vec{r}_j)||\right)$ depends on the relative distance $r$ between the pair of molecules $(D^*_{\alpha},A_{j})$, and is obtained from FRET theory \cite{forster1960transfer,doi:10.1021/jp900708f} as 
\begin{eqnarray}
k(r) = \Gamma_D \left( \frac{R_0}{r} \right)^6
\label{Reac4} \, .
\end{eqnarray}
For simplicity reasons, in Eq.\ref{Reac4}, we did not take into account the full dependence of the rate $k(r,\theta)$ with the relative orientation (through the angle $\theta$) of the molecular dipoles  \cite{doi:10.1063/1.4861695}.  
This dependence and its impact on the RET dynamics, is beyond the scope of the present paper.
It was shown to lead to some corrections of the FRET theory for intermolecular distances larger than $10 \mbox{ nm}$ \cite{doi:10.1063/1.4861695}.
We could in principle add such effects to our model: however, we chose to treat only one level of complexity, namely the one arising from \textit{the dependence of RET with intermolecular distances and spatial inhomogeneities of the molecules' location in space.}
This implicitly assumes that an averaging on molecular dipole orientations has been performed in Eq.\ref{Reac4}, that restores the rotational invariance of the microscopic rate $k(r)$.  
%

%
%
\subsection{RET master equation}
\label{CME}
We introduce the \textit{microstates of the molecular ensemble} as the list of excitation-states for each molecule $a \equiv \left\lbrace \left\lbrace \sigma_\alpha \right\rbrace_{\alpha\in\lbrack |1,N_D| \rbrack},  \left\lbrace \sigma_j \right\rbrace_{j\in\lbrack |1,N_A| \rbrack} \right\rbrace$,
in which $\sigma_\alpha=1(0)$ if the donor molecule $\alpha$ is in excited (ground) state $D^{*}_\alpha(D_\alpha)$, and $\sigma_j=1(0)$ if the acceptor molecule $j$ is in its excited (ground) state $A^{*}_j(A_j)$.
Under the influence of the elementary reaction steps given by Eqs.\ref{Reac1},\ref{Reac2},\ref{Reac3}, the microstates $a$ undergo a Markov stochastic process, that is described by the following microscopic \textit{master equation} (ME) \cite{van1992stochastic}
\begin{eqnarray}
\dot{P}_a(t) &=& \sum_b \left\lbrace \Gamma_{ab} P_b(t) - \Gamma_{ba} P_a(t) \right\rbrace
\label{CME1} \, ,
\end{eqnarray}
with $P_a(t)$ the probability of occupying the microstate $a$ at time $t$ and $\Gamma_{ba}$ \textcolor{black}{the incoherent rate between the microstates for the transition $a \rightarrow b$.}
%
%
The latter rate is $\Gamma_{ba}=\Gamma_{D(A)}$ for a transition involving a spontaneous photon emission event as described by Eq.\ref{Reac1} (Eq.\ref{Reac2}). 
Transitions involving a RET process are given by Eq.\ref{Reac3}, and contribute with a rate
$\Gamma_{ba}=k_{\alpha j}$.
The initial condition at time $t=0$ is fixed to be the microstate $a_{\rm{in}}\equiv \left\lbrace \left\lbrace \sigma_\alpha=1 \right\rbrace_{\alpha\in\lbrack |1,N_D| \rbrack},  \left\lbrace \sigma_j=0 \right\rbrace_{j\in\lbrack |1,N_A| \rbrack} \right\rbrace$ in which all donor molecules are excited and acceptor ones are in their ground-state, so that  
\begin{eqnarray}
P_a(0) &=& \delta_{a,a_{\rm{in}}}
\label{CME2} \, .
\end{eqnarray}
The conjunction of Eq.\ref{CME1} and Eq.\ref{CME2} provides a complete description (including spatial inhomogeneity) of the RET stochastic dynamics from the initial (fully excited) microstate $a_{\rm{in}}$ to the final (fully relaxed) microstate $a_{\rm{f}}\equiv \left\lbrace \left\lbrace \sigma_\alpha=0 \right\rbrace_{\alpha\in\lbrack |1,N_D| \rbrack},  \left\lbrace \sigma_j=0 \right\rbrace_{j\in\lbrack |1,N_A| \rbrack} \right\rbrace$, for which all molecules have decayed to their ground-state. 
We note the similarity in our modelling of the RET process, with the approach of Paillotin \textit{et al.} \cite{PAILLOTIN1979513} developed for analyzing the impact of exciton-exciton annihilation processes on transient fluorescence spectra in photosynthetic systems.
%

%
%
\subsection{Kinetic Monte-Carlo approach}
\label{Kin}
%
%
In principle, the solution of the ME enables to find the average number of excited donor molecules $\left\langle N_{D^*}(t) \right\rangle$ at each time $t$, as
\begin{eqnarray}
\left\langle N_{D^*}(t) \right\rangle &=& \sum_a \left( N_{D^*} \right)_a P_a(t)
\label{Kin1} \, ,
\end{eqnarray}
with $\left( N_{D^*} \right)_a \equiv \sum_{\alpha\in\lbrack |1,N_D|\rbrack} \sigma_{\alpha}$, the number of excited donor molecules in the configuration $a$.
A similar expression is obtained for the average number of acceptor molecules in their ground-state $\left\langle N_{A}(t) \right\rangle$, at time $t$.
However, due to the large number $N$ of molecules, and resulting exponentially larger number of possible microstates for the system (scaling with $2^N$), \textit{it is in practise very hard to solve the ME, despite the fact it is a linear equation.}
%

%
%
We thus resort to solve numerically the ME, using a kinetic Monte Carlo (MC) \cite{GILLESPIE1976403,doi:10.1063/1.461138} approach.
The main idea of it, is to discretize the time-window into small time-steps $\delta t$.
At each time-step, one either updates randomly the $a$ microstate of the system to the $b$ one with probability $\propto \Gamma_{ba} \delta t$, or leaves the $a$-state unchanged. 
The number of excited donors and acceptors in the ground-state is then counted and stored.
%
%
%
\textcolor{black}{The final observable values $\left\langle N_{D^*}(t) \right\rangle$ and $\left\langle N_{A}(t) \right\rangle$ are obtained after averaging on a large-enough number $N_{traj}$ of stochastic trajectories of the system connecting the initial to the final state.
The chosen value of $N_{traj}=10^5$ is sufficient to reach a relative statistical uncertainty of $\approx 1/\sqrt{N_{traj}} \approx 0.3\%$ for the MC calculation (see Appendix \ref{Unc_MC} for a more detailed discussion).
}
%

\section{Kinetic equations}
\label{Sec2}
%
%
%
\subsection{Derivation of the kinetic equations}
\label{KinEq}
\begin{figure}[tbh]
\includegraphics[width=1.0\linewidth, angle=-0]{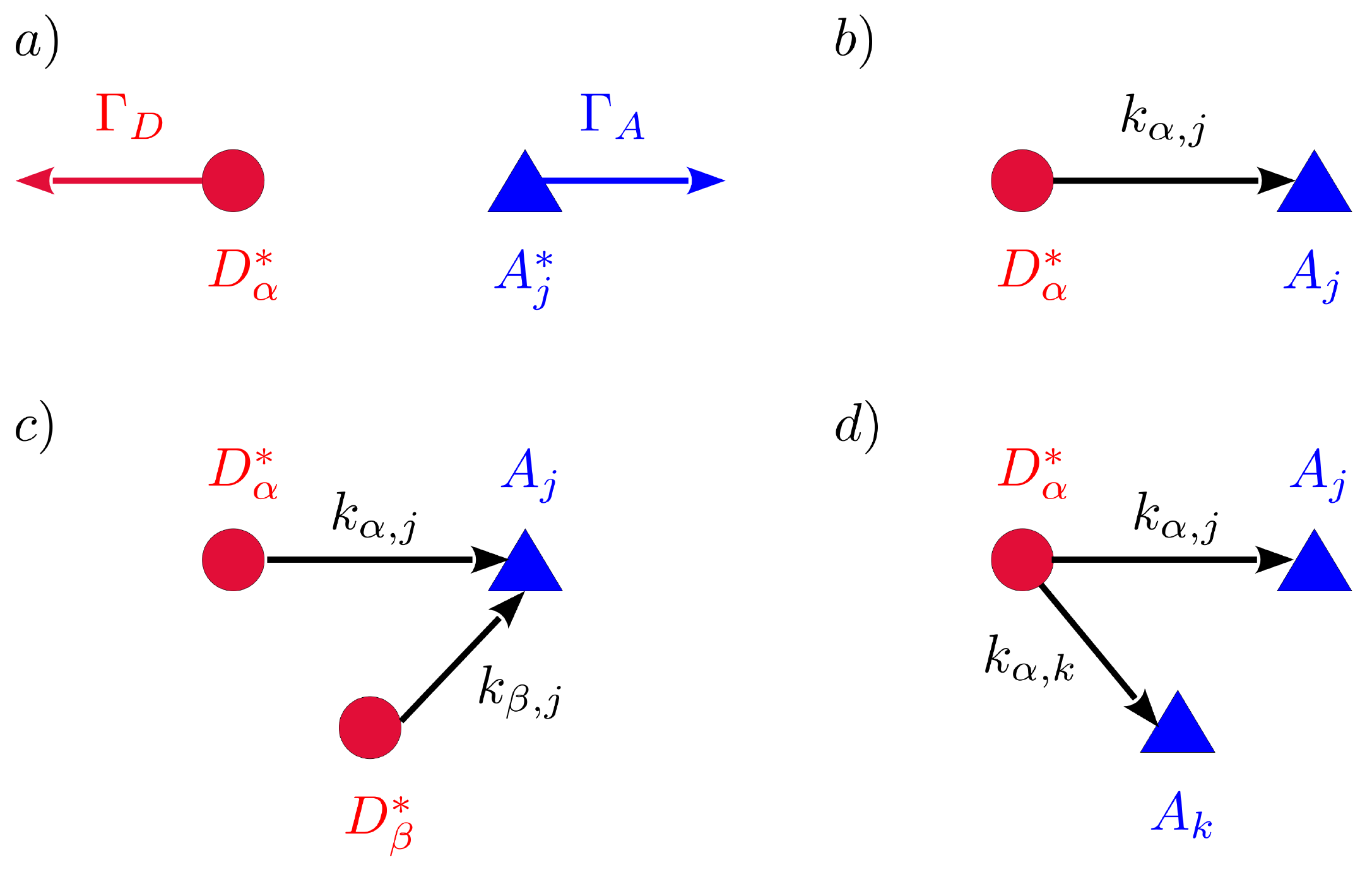}
\caption{
a) One-molecule fluorescence processes involving spontaneous emission of a photon with rate $\Gamma_{D(A)}$ from an excited $D^*_\alpha(A^*_j)$ molecule located at site $\alpha(j)$.
b) Two-molecule RET process with rate $k_{\alpha,j}$ between $D^*_\alpha$ and $A_j$.
c) Three-molecule competition between two RET processes with rates $k_{\alpha,j}$ and $k_{\beta,j}$ involving the molecules in state $D^*_\alpha$, $D^*_\beta$ and $A_j$.
d) Three-molecule competition between two RET process with rates $k_{\alpha,j}$ and $k_{\alpha,k}$ involving the molecules in state $D^*_\alpha$, $A_k$ and $A_j$.
}
\label{fig:Fig2}
\end{figure}
%
%
%
In this section, we derive exact kinetic equations from Eq.\ref{CME1}, that describe the time-dependence of the average populations of excited donors $\left\langle N_{D^*}(t) \right\rangle$ and acceptors in the ground-state $\left\langle N_{A}(t) \right\rangle$.
Those average populations are given by
\begin{eqnarray}
\left\langle N_{D^*}(t) \right\rangle &=& \sum_{\alpha=1}^{N_D}\left\langle \sigma_\alpha(t) \right\rangle
\label{KinEq1} \, , \\
\left\langle N_{A}(t) \right\rangle &=& \sum_{j=1}^{N_A}\left\langle \bsigma_j(t) \right\rangle
\label{KinEq2} \, , 
\end{eqnarray}
with $\bsigma_j=1-\sigma_j$, and 
\begin{eqnarray}
\left\langle \sigma_\alpha(t) \right\rangle \equiv \sum_{\sigma_\alpha=0,1} \sigma_\alpha \Pi^{(1)}_{\sigma_\alpha}(t)
\label{KinEq2bis} \, ,
\end{eqnarray}
the average excitation state of the donor $\alpha$.
In Eq.\ref{KinEq2bis}, we introduced $\Pi^{(1)}_{\sigma_\alpha}(t)$, the "one-molecule" probability distribution that the donor $\alpha$ is in excitation state $\sigma_\alpha$ at time $t$.
A similar definition is introduced for $\left\langle \bsigma_j(t) \right\rangle$.
%

%
%
From Eqs.\ref{CME1},\ref{KinEq2bis}, we can derive exact kinetic equations (KE)
for the time-dependent fields $\left\langle \sigma_\alpha(t) \right\rangle$ and $\left\langle \bsigma_j(t) \right\rangle$ involved in the calculation of the average populations
%
%
\begin{eqnarray}
\frac{d}{dt}\left\langle \sigma_\alpha \right\rangle &=& -\Gamma_D \left\langle \sigma_\alpha \right\rangle - \sum_{j=1}^{N_A} k_{\alpha,j} \left\langle \sigma_\alpha \bsigma_j \right\rangle
\label{KinEq3} \, , \\
\frac{d}{dt}\left\langle \bsigma_j \right\rangle &=& \Gamma_A \left\langle \sigma_j \right\rangle - \sum_{\alpha=1}^{N_D} k_{\alpha,j} \left\langle \sigma_\alpha \bsigma_j \right\rangle
\label{KinEq4} \, ,
\end{eqnarray}
with
\begin{eqnarray}
\left\langle \sigma_\alpha(t)\bsigma_j(t) \right\rangle \equiv \sum_{\sigma_\alpha=0,1}\sum_{\sigma_j=0,1} \sigma_\alpha \bsigma_j \Pi^{(2)}_{\sigma_\alpha,\sigma_j}(t)
\label{KinEq4bis} \, ,
\end{eqnarray}
the correlation function between the states of occupancy of the donor molecule $\alpha$ and of the acceptor molecule $j$.
The latter involves the joint probability distribution $\Pi^{(2)}_{\sigma_\alpha,\sigma_j}(t)$ that the donor molecule $\alpha$ is in state $\sigma_\alpha$ and the acceptor molecule $j$ is in state $\sigma_j$.
%

%
%
Eqs.\ref{KinEq3},\ref{KinEq4} have a clear physical interpretation. 
For instance, in Eq.\ref{KinEq3}, the first term on the right-hand-side (rhs) stands for the decay channel from the excited-state of the donor $\alpha$ by fluorescence with rate $\Gamma_D$ (see Fig.~\ref{fig:Fig2}-a)). 
The second term on the rhs, is related to the RET reaction with rate $k_{\alpha,j}$, that naturally correlates the excited-state of the donor $\alpha$ to the ground-state of the acceptor $j$ (see Fig.~\ref{fig:Fig2}-b)).
The KE have to be complemented with a set of initial conditions
\begin{eqnarray}
&&\left\langle \sigma_\alpha(0) \right\rangle = \left\langle \bsigma_j(0) \right\rangle = 1
\label{KinEq5} \, , \\
&&\left\langle \sigma_\alpha(0) \bsigma_j(0) \right\rangle = 1
\label{KinEq6} \, .
\end{eqnarray}
As expected, the KE fulfil at each time $t$ the laws of mass-conservation in the chemical process 
\begin{eqnarray}
\left\langle N_{D^*}(t) \right\rangle  + \left\langle N_{D}(t) \right\rangle &=& N_D 
\label{KinEq7} \, , \\
\left\langle N_{A^*}(t) \right\rangle  + \left\langle N_{A}(t) \right\rangle &=& N_A 
\label{KinEq8} \, .
\end{eqnarray}
%

%
%
\subsection{Evolution of correlation functions}
\label{CorrEq}
%
%
Despite being exact, the KE do not constitute a closed-system of equations. 
Indeed, one has to complement Eqs.\ref{KinEq3},\ref{KinEq4} with additional equations providing the time-evolution of the still unknown correlation functions $\left\langle \sigma_\alpha(t)\bsigma_j(t) \right\rangle$.
As in Sec.\ref{KinEq}, we derive the following equations for the correlation functions
\begin{eqnarray}
&&\frac{d}{dt}\left\langle \sigma_\alpha\bsigma_j \right\rangle = \Gamma_A \left\langle \sigma_\alpha\sigma_j\right\rangle -\left( \Gamma_D + k_{\alpha,j} \right) \left\langle \sigma_\alpha\bsigma_j \right\rangle \nonumber \\
&-& \sum_{\beta=1,\beta\neq \alpha}^{N_D} k_{\beta,j} \left\langle \sigma_\alpha\sigma_\beta\bsigma_j\right\rangle - \sum_{k=1,k\neq j}^{N_A} k_{\alpha,k} \left\langle \sigma_\alpha\bsigma_k\bsigma_j\right\rangle
\nonumber \, , \\
\label{CorrEq1}  
\end{eqnarray}
with $\left\langle \sigma_\alpha(t)\sigma_j(t)\right\rangle = \left\langle \sigma_\alpha(t)\right\rangle - \left\langle \sigma_\alpha(t)\bsigma_j(t)\right\rangle$ and the initial conditions
\begin{eqnarray}
\left\langle \sigma_\alpha(0)\sigma_\beta(0)\bsigma_j(0)\right\rangle = \left\langle \sigma_\alpha(0)\bsigma_k(0)\bsigma_j(0)\right\rangle = 1
\label{CorrEq2} \, . 
\end{eqnarray}
The interpretation of Eq.\ref{CorrEq1} is the following. 
The first three terms in the rhs come from "one-molecule" fluorescence processes and "two-molecules" RET process. 
The last two terms on the rhs involve "three-molecules" correlation functions $\left\langle \sigma_\alpha(t)\sigma_\beta(t)\bsigma_j(t)\right\rangle$ and $\left\langle \sigma_\alpha(t)\bsigma_k(t)\bsigma_j(t)\right\rangle$.
We interpret those new two terms as coming from a competition mechanism in the RET process between two excited donors for the same acceptor in its ground-state (see Fig.~\ref{fig:Fig2}-c), or between one excited donor and two different acceptors in their ground-states (see Fig.~\ref{fig:Fig2}-d).
%

%
%
Similarly to the KE, Eq.\ref{CorrEq1} is not closed. 
Thus, it has to be supplemented by another equation describing the dynamics of the unknown three-molecules correlation functions.
\textit{This derivation can be proceeded further by iteration, producing a hierarchy of differential equations, coupling the dynamics of the populations to all higher-order correlation functions of the molecular states, making the complete KE as difficult to solve as the initial ME}.
%

%
%
\section{Mean-field approximation}
\label{Mean-field}
%
%
%
\subsection{Macroscopic limit}
\label{Macro}
%
%
%
In this section, we aim at simplifying the KE, by using a closure assumption of Eqs.\ref{KinEq3},\ref{KinEq4} which reproduces well the macroscopic limit obtained in the limit of large number of molecules $N \gg 1$, or large sample volume $\Omega$ \textcolor{black}{(given by the sample surface $S$ in dimension two).}
We make \textit{a first assumption $(A1)$ that in the large-N limit, the two-molecules joint-distribution function $\overline{\Pi}^{(2)}_{\sigma_\alpha,\sigma_j}(t)$ in Eq.\ref{KinEq4bis} factorizes}, namely that
\begin{eqnarray}
\overline{\Pi}^{(2)}_{\sigma_\alpha,\sigma_j}(t) \approx \overline{\Pi}^{(1)}_{\sigma_\alpha}(t) \overline{\Pi}^{(1)}_{\sigma_j}(t)
\label{Marcro1} \, ,
\end{eqnarray}
where $\overline{A}$ means the coarsed-grained (spatially averaged) value associated to any quantity 
$A$.
Using this approximation $(A1)$, we can simplify the KE as
\begin{eqnarray}
\frac{d}{dt}\overline{\left\langle \sigma_\alpha \right\rangle} &=& -\Gamma_D \overline{\left\langle \sigma_\alpha \right\rangle} - \sum_{j=1}^{N_A} k_{\alpha,j} \overline{\left\langle \sigma_\alpha\right\rangle} \mbox{ }\mbox{ } \overline{\left\langle\bsigma_j \right\rangle}
\label{Macro2} \, , \\
\frac{d}{dt}\overline{\left\langle \bsigma_j \right\rangle} &=& \Gamma_A \overline{\left\langle \sigma_j \right\rangle}
 - \sum_{\alpha=1}^{N_D} k_{\alpha,j} \overline{\left\langle \sigma_\alpha\right\rangle} \mbox{ }\mbox{ } \overline{\left\langle\bsigma_j \right\rangle}
\label{Macro3} \, .
\end{eqnarray}
Those macroscopic KE are closed for the fields $\left\lbrace\overline{\left\langle \sigma_\alpha \right\rangle}\right\rbrace_\alpha$ and $\left\lbrace\overline{\left\langle \bsigma_j \right\rangle}\right\rbrace_j$, and constitute a system of non-linear mean-field equations, that take into account spatial inhomogeneity of the distribution of molecules in the sample.
Such KE can in principle be solved with initial conditions given by Eq.\ref{KinEq5}, although, due to the large number of unknowns and the non-linearity of the equations, this remains a heavy and difficult task.
%

%
%
We now make \textit{the last assumption $(A2)$, that after spatial coarse-graining, the molecular system becomes spatially homogeneous in the large-N limit}, namely that translational invariance is restored for the mean-fields 
\begin{eqnarray}
\overline{\left\langle \sigma_\alpha (t) \right\rangle} &\approx& \phi^{\rm{(mf)}}_{D^*}(t) = \frac{\left\langle N_{D^*}(t) \right\rangle}{N_D}  
\label{Macro4} \, , \\
\overline{\left\langle \bsigma_j (t) \right\rangle} &\approx& \phi^{\rm{(mf)}}_{A}(t) = \frac{\left\langle N_{A}(t) \right\rangle}{N_A}  
\label{Macro4bis} \, , 
\end{eqnarray}
Using Eqs.\ref{Macro4},\ref{Macro4bis} inside Eqs.\ref{Macro2},\ref{Macro3}, we derive simpler KE for the populations in the macroscopic limit
\begin{eqnarray}
\frac{d}{dt}\left\langle N_{D^*}(t) \right\rangle &=& -\Gamma_D \left\langle N_{D^*}(t) \right\rangle - \overline{k}\left\langle N_{D^*}(t) \right\rangle \left\langle N_{A}(t) \right\rangle
\nonumber \, , \\
\label{Macro5} \\
\frac{d}{dt}  \left\langle N_{A}(t) \right\rangle &=& \Gamma_A  \left\langle N_{A^*}(t) \right\rangle 
- \overline{k}\left\langle N_{D^*}(t) \right\rangle \left\langle N_{A}(t) \right\rangle 
\nonumber \, , \\
\label{Macro6} 
\end{eqnarray}
with the mean-field rate of RET
\begin{eqnarray}
\overline{k} &=& \frac{1}{N_D N_A} \sum_{\alpha=1}^{N_D}\sum_{j=1}^{N_A} k_{\alpha,j}
\label{Macro7} \, .
\end{eqnarray}
Eqs.\ref{Macro5},\ref{Macro6} are the main results of this section. 
They could have been guessed directly from the elementary steps of the RET reaction (see Eq.\ref{Reac1},\ref{Reac2},\ref{Reac3}), since they simply express that the fluorescence process has a kinetics of order one while the RET process has a kinetics of order two in the reactant concentrations. 
However, our theory provides \textit{a solid approach to recover this macroscopic limit, to compute the coarsed grained reaction rate $\overline{k}$, as given by Eq.\ref{Macro7}, and to investigate effects of spatial inhomogeneities and fluctuations in mesoscopic samples beyond standard macroscopic KE}.
%

%
%
In the case of an equimolar sample of donors and acceptors ($N_D=N_A=N/2$) with the same fluorescence rate ($\Gamma_D=\Gamma_A\equiv\Gamma$), the previous macroscopic KE for the populations can be solved exactly, leading to 
\begin{eqnarray}
\left\langle N_{D^*}(t) \right\rangle^{\rm{(mf)}} &=& \frac{N}{2} \Xi(t) \left\lbrace 1+\tilde{k}\int_0^t d\tau \Xi(\tau) \right\rbrace^{-1}
\label{Macro8} \, , \\
\left\langle N_{A}(t) \right\rangle^{\rm{(mf)}} &=& \frac{N}{2}\left( 1 - e^{-\Gamma t} \right) + \left\langle N_{D^*}(t) \right\rangle^{\rm{(mf)}}
\label{Macro9} \, , 
\end{eqnarray}
with 
\begin{eqnarray}
\Xi(t) &=& \exp\left\lbrace
-\left( \Gamma + \tilde{k} \right) t 
+ 
\frac{\tilde{k}}{\Gamma} \left(
1 - e^{-\Gamma t}
\right)
\right\rbrace
\label{Macro10} \, , \\
\tilde{k} &=& \lim_{N\rightarrow +\infty} \frac{N}{2} \overline{k} 
\equiv 
\lim_{N\rightarrow +\infty} \frac{2}{N} \sum_{\alpha=1}^{N_D}\sum_{j=1}^{N_A} k_{\alpha,j}
\label{Macro11} \, .
\end{eqnarray}
The macroscopic rate of RET $\tilde{k}$ is obtained quantitatively in Eq.\ref{Macro11}. 
In the large-N limit, $\overline{k}\rightarrow 0$ while $\tilde{k}$ remains finite. 
%

%
%
\subsection{Mean-field rate of RET}
\label{FRETmf}
\begin{figure}[tbh]
\includegraphics[width=1.0\linewidth, angle=-0]{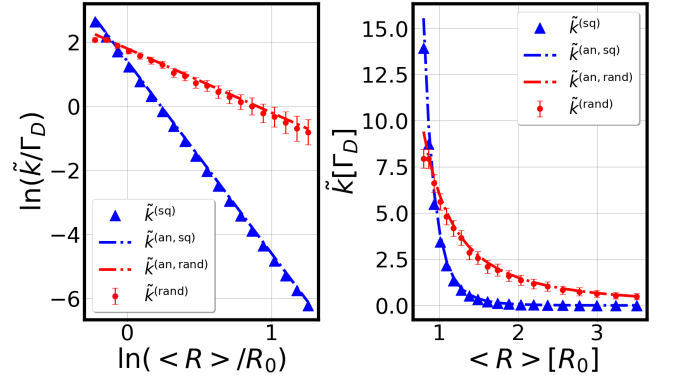}
\caption{
\textcolor{black}{
Left: Numerical computation of the macroscopic mean-field rate of RET $\tilde{k}^{\rm{(rand)}}$ in log-scale (red points) from Eq.\ref{Macro11}, as a function of the average distance $<R>$ between individual excited donors and their nearest-neighbor acceptor molecules.
An average is performed on $1000$ configurations of disorder, for various random distributions of molecules on the square network shown in Fig.~\ref{fig:Fig1}-Right.}
\textcolor{black}{
The error bars represent one standard deviation, illustrating the uncertainty due to fluctuations in the spatial location of molecules that varies from sample to sample.
The dash-dotted red curve is obtained from the analytical rate $\tilde{k}^{\rm{(an,rand)}}$ in Eq.\ref{PerfDis2}.}
\textcolor{black}{
The same calculation for the rate $\tilde{k}^{\rm{(sq)}}$ is shown (blue triangles) for the case of one perfectly ordered square network of molecules shown in Fig.~\ref{fig:Fig1}-Left.
The dash-dotted blue curve is given by the analytical rate $\tilde{k}^{\rm{(an,sq)}}$ in Eq.\ref{PerfCrys1}.}
%
%
%
\textcolor{black}{
Right: Same plots but in linear scale. 
}
Parameters: $N_D=N_A=N/2=50$, and $R_c = 0.8R_0$.
}
\label{fig:Fig3}
\end{figure}
%
%
%
In this section, we provide explicit expressions of the macroscopic mean-field rate of RET $\tilde{k}$ for different examples of bi-dimensional samples. 
To this purpose, we first rewrite Eq.\ref{Macro11} in integral form
\begin{eqnarray}
\tilde{k} &=& \frac{2}{N} \int d^2\vec{r} \int d^2\vec{r'} \overline{\rho_{DA}}(\vec{r},\vec{r'})
k\left( || \vec{r}-\vec{r'} || \right)
\label{FRETmf1} \, ,
\end{eqnarray}
with $k\left( || \vec{r}-\vec{r'} || \right)$ given by Eq.\ref{Reac4}. 
The spatially averaged density of pairs of donor and acceptor molecules $\overline{\rho_{DA}}(\vec{r},\vec{r'})$ is defined as
\begin{eqnarray}
 \overline{\rho_{DA}}(\vec{r},\vec{r'}) &=& \overline{\rho_{D}(\vec{r})\rho_{A}(\vec{r'})}
\label{FRETmf2} \, ,
\end{eqnarray}
with $\rho_{D}(\vec{r}) = \sum_{\alpha=1}^{N_D} \delta^2\left( \vec{r} - \vec{r}_\alpha \right)$ and $\rho_{A}(\vec{r'}) = \sum_{j=1}^{N_A} \delta^2\left( \vec{r'} - \vec{r}_j \right)$. 
%
%
The assumption $(A2)$ of translational invariance implies that $ \overline{\rho_{DA}}(\vec{r},\vec{r'}) \equiv  \overline{\rho_{DA}}(\vec{r}-\vec{r'})$, so that the mean-field rate becomes in polar coordinates
\begin{eqnarray}
\tilde{k} &=& \frac{2 S}{N} \Gamma_D R^6_0 \int_{R_c}^{R} dr \int_0^{2\pi}  d\theta\frac{\overline{\rho_{DA}}(r,\theta)}{r^5}
\label{FRETmf3} \, ,
\end{eqnarray}
\textcolor{black}{with $S=\pi R^2$ the surface of the sample of radius $R$, and $R_c$ the same cutoff length used in Fig.~\ref{fig:Fig1}-Right defined as the minimum distance between two molecules}.
This expression of $\tilde{k}$ is general and independent of the chosen network of molecules. 
It connects the effective rate of RET in our simple model, to the actual spatial distribution of molecular pairs of donors and acceptors in the macroscopic sample. 
%

%
%
\subsubsection{Disordered square lattice}
\label{PerfDis}
%
%
%
\textcolor{black}{We suppose here that the donor and acceptor molecules are randomly distributed forming a random lattice (see Fig.1-Right), namely they form an homogeneous sample of reacting molecules.} 
%
%
The assumptions of factorization of the "two-molecules" correlation functions (assumption $(A1)$) and of translational invariance (assumption $(A2)$) after spatial coarse graining, imply that the averaged spatial density of pairs in Eq.\ref{FRETmf2} simplifies to
\begin{eqnarray}
 \overline{\rho_{DA}}(\vec{r},\vec{r'}) &\approx & \overline{\rho}_{D} \overline{\rho}_{A}
 \label{PerfDis1} \, ,
\end{eqnarray}
with $\overline{\rho}_{D} = \overline{\rho}_{A} \approx N/2S$, the equal concentration of donor and acceptor molecules.
This provides a simpler and explicit expression of $\tilde{k}$ in Eq.\ref{FRETmf3}, which is valid in the limit of large number of molecules $N\gg 1$ or of large sample surface $S \gg R^2_c, R^2_0$.
We obtain
\begin{eqnarray}
\tilde{k}^{(\rm{an,rand})} &\approx & k_0 \left( \frac{R_0}{2 \left\langle R \right\rangle} \right)^2
\label{PerfDis2} \, ,
\end{eqnarray}
with \textcolor{black}{$k_0=\Gamma_D\left(R_0/R_c \right)^4$} and $\left\langle R \right\rangle = \sqrt{S/\pi N}$ the average distance between an individual excited-donor and its nearest-neighbor acceptor (proportional to the inverse square root of the molecular concentration). 
%
%
%
Eq.\ref{PerfDis2} is a scaling-law $\tilde{k} \propto \left\langle R \right\rangle^{-2}$ for the macroscopic mean-field rate of RET, in a disordered sample.
This is a surprising result, since one would have expected a scaling $\tilde{k} \propto \left\langle R \right\rangle^{-6}$ by simple inference from the microscopic rate in Eq.\ref{Reac4}.
This interesting effect comes from disorder in the bi-dimensional sample, namely \textit{the average rate of RET from a given donor molecule results from spatial averaging of the contribution of acceptor molecules around it, and thus scales with the molecular concentration.} 
%
%
%
\textcolor{black}{
We show on Fig.~\ref{fig:Fig3}-Left (red points), the dependence of $\tilde{k}^{(\rm{rand})}$ in units of $\Gamma_D$, as a function of $<R>/R_0$ in logarithmic scale. 
This curve is obtained from a numerical evaluation of Eq.\ref{Macro11}, for a random lattice containing $N_D=N_A=50$ molecules, after averaging on $1000$ configurations of disorder.
The same plot is shown in linear scale in Fig.~\ref{fig:Fig3}-Right (red points).
This calculation compares quantitatively very well to the predicted analytical values of $\tilde{k}^{(\rm{an,rand})}$ given by Eq.\ref{PerfDis2} (see dash-dotted red curves).
For completeness, we show error bars representing one standard deviation (see also Appendix \ref{Unc_k_mf}), and illustrating the uncertainty due to fluctuations in the computed values of $\tilde{k}^{(\rm{rand})}$ when considering different disorder-configurations of the random network of molecules.}
%
%
%

%
%
\subsubsection{Perfectly ordered square lattice}
\label{PerfCrys}
%
%
In contrast to the previous example, we consider now the opposite case of a distribution of donor and acceptor molecules located at the vertex of a perfectly ordered square lattice (shown in Fig.~\ref{fig:Fig1}-Left). 
The unit cell contains four molecules, with alternating $D$ and $A$ molecules being separated by the D-A distance $a$ (the lattice parameter thus being $2a$).
The macroscopic mean-field rate of RET can be evaluated analytically from Eq.\ref{Macro11} as
\begin{eqnarray}
\tilde{k}^{(\rm{an,sq})} &=& 2 \Psi \Gamma_D \left( \frac{R_0}{2a} \right)^6 
\label{PerfCrys1} \, ,
\end{eqnarray}
with $\Psi=\sum_{(m,n)\in\mathbb{Z}^2}\left\lbrace \left( n + 1/2 \right)^2 + m^2 \right\rbrace^{-3}\approx 130,45$ a pure number given by an absolutely convergent double series characterizing the bi-dimensional square lattice.
%

%
%
In contrast to Eq.\ref{PerfDis2}, the effective rate of RET scales here as $\tilde{k} \propto \left\langle R \right\rangle^{-6}$ (since $\left\langle R \right\rangle=a$) which matches with the expected scaling-law given by the microscopic theory of FRET in Eq.\ref{Reac4}. 
This is due to the fact that \textit{our square lattice is an ordered lattice of pairs of D-A molecules repeating regularly, each $D$ molecule having four nearest-neighbours $A$ molecules.} 
%
%
%
\textcolor{black}{
We show on Fig.~\ref{fig:Fig3} (blue triangles), the corresponding rate $\tilde{k}^{(\rm{sq})}$, as a function of $<R>/R_0$ in logarithmic and linear scale, computed numerically from Eq.\ref{Macro11}, in the case of a regular square lattice containing $N_D=N_A=50$ molecules.
A very good agreement is obtained with the analytical result provided by $\tilde{k}^{(\rm{an,sq})}$
in Eq.\ref{PerfCrys1} (see dash-dotted blue curve).
%
%
Contrary to the case of the disordered lattice, we do not show here any error bars, since there is only one spatial configuration of the ordered regular lattice (see Fig.~\ref{fig:Fig1}-Left) taken into account in the calculation of $\tilde{k}^{(\rm{sq})}$.
}
%
%
%

\subsection{Fluctuations and Collective effects}
\label{Fluctuations}
In order to understand better the range of validity of the mean-field approximation with respect to the numerically exact MC calculations, we investigate in this section the role of fluctuations in the RET dynamics. 
In the case of an arbitrary lattice, this is a rather formidable task.
We thus restrict in our derivation to the case of an homogeneous system, for which we are able to investigate the role of fluctuations developing out-of-equilibrium \cite{Fluctuations_Nonequilibrium_Systems,nitzan1974comment}.
Following the approach of van Kampen \cite{van1992stochastic}, we derive \textit{an asymptotic expansion of the chemical ME in the limit of large number $N$ of molecules or equivalently of large volume $\Omega\equiv S$ of the sample}.
The intermediate calculations are analytical but lengthy, and we thus only provide here the main steps of the derivation and the final results.
For this purpose, we introduce the notation
\begin{eqnarray}
N_{D^*}(t) &=& \textcolor{black}{N_D} \phi_{D^*}^{\rm{(mf)}}(t) + \textcolor{black}{\sqrt{N_D}} \xi_{D}(t)
\label{CME_Hom_1} \, , \\
N_{A^*}(t) &=& \textcolor{black}{N_A} \phi_{A^*}^{\rm{(mf)}}(t) + \textcolor{black}{\sqrt{N_A}} \eta_{A}(t)
\label{CME_Hom_2} \, , 
\end{eqnarray}
%
%
%
%
%
%
with $\phi_{D^*}^{\rm{(mf)}}(t)$ and $\phi_{A^*}^{\rm{(mf)}}(t)$, the unknown (at this stage) normalized average populations of excited donors and acceptors respectively, and $\xi_{D}(t)$ and $ \eta_{A}(t)$ the associated fluctuations with respect to those averages.
The chemical ME can be expanded systematically in the small parameter $\Omega^{-1}$ \cite{van1992stochastic}.
At leading-order $\rm{o}(\Omega^{1/2})$, we recover the macroscopic mean-field equations provided by Eqs.\ref{Macro5},\ref{Macro6}, namely that $\phi_{D^*}^{\rm{(mf)}}(t) \equiv \left\langle N_{D^*}(t)\right\rangle^{\rm{(mf)}}/N_D$ and $\phi_{A^*}^{\rm{(mf)}}(t) \equiv\left\langle N_{A^*}(t)\right\rangle^{\rm{(mf)}}/N_A$. 
The term of order $\rm{o}(\Omega^0)$ in the expansion provides a multivariate Fokker-Planck equation describing Gaussian fluctuations $\xi_{D}(t)$ and $ \eta_{A}(t)$ around the mean-field solutions $\phi_{D^*}^{\rm{(mf)}}(t)$ and $\phi_{A^*}^{\rm{(mf)}}(t)$. 
We derive from it a closed system of linear, time-dependent equations describing the evolution of the following correlation functions $\left\langle \xi_D(t) \eta_A(t) \right\rangle $, $ \left\langle \xi^2_D(t) \right\rangle$ and $ \left\langle \eta^2_A(t) \right\rangle$.
We obtain
\begin{eqnarray}
\frac{d}{dt}\left\langle \xi_D \eta_A \right\rangle &=& - \left\lbrack \Gamma_A+\Gamma_D + \tilde{k} \left( \phi^{\rm{(mf)}}_{A} + \phi^{\rm{(mf)}}_{D^*} \right) \right\rbrack \left\langle \xi_D \eta_A \right\rangle
\nonumber \\
&+& \tilde{k} \lbrack \phi^{\rm{(mf)}}_{A} \left\langle \xi^2_D \right\rangle + \phi^{\rm{(mf)}}_{D^*} \left\langle \eta^2_A \right\rangle \rbrack - \tilde{k} \phi^{\rm{(mf)}}_{A} \phi^{\rm{(mf)}}_{D^*}
\nonumber \, , \\
\label{CME_Hom_3} \\
\frac{d}{dt}\left\langle \xi^2_D\right\rangle &=& - 2 \left( \Gamma_D + \tilde{k} \phi^{\rm{(mf)}}_{A} \right) \left\langle \xi_D^2 \right\rangle
\nonumber \\
&+& 2\tilde{k} \phi^{\rm{(mf)}}_{D^*} \left\langle \xi_D \eta_A \right\rangle 
+ \left(  \Gamma_D + \tilde{k} \phi^{\rm{(mf)}}_{A} \right) \phi^{\rm{(mf)}}_{D^*} 
\nonumber \, , \\
\label{CME_Hom_4} \\
\frac{d}{dt}\left\langle \eta^2_A\right\rangle &=& - 2 \left( \Gamma_A + \tilde{k} \phi^{\rm{(mf)}}_{D^*} \right) \left\langle \eta_A^2 \right\rangle
\nonumber \\
&+& 2\tilde{k} \phi^{\rm{(mf)}}_{A} \left\langle \xi_D \eta_A \right\rangle 
+ \Gamma_A \phi^{\rm{(mf)}}_{A^*} + \tilde{k} \phi^{\rm{(mf)}}_{A}\phi^{\rm{(mf)}}_{D^*} 
\nonumber \, , \\
\label{CME_Hom_5}  
\end{eqnarray}
with $\phi^{\rm{(mf)}}_{A}(t)=1-\phi^{\rm{(mf)}}_{A^*}(t)$ and the initial conditions $\left\langle \xi_D(0) \eta_A(0) \right\rangle = \left\langle \xi^2_D(0) \right\rangle = \left\langle \eta^2_A(0) \right\rangle = 0$.
Those equations are the main results of this section. 
Despite their complexity, they have a simple physical interpretation.
For instance in the first term of Eq.\ref{CME_Hom_3}, the correlation function $\left\langle \xi_D(t) \eta_A(t) \right\rangle$ decays in time (regression of fluctuations) due to the decay of each $\xi_D(t)$ and $\eta_A(t)$ with the respective rates $\Gamma_D + \tilde{k} \phi^{\rm{(mf)}}_{A}(t)$ and $\Gamma_A + \tilde{k} \phi^{\rm{(mf)}}_{D^*}(t)$.
\textit{The latter rates are explicitely time-dependent due to the non-linearity of the RET process.}
The second term of Eq.\ref{CME_Hom_3} corresponds to an increase of $\left\langle \xi_D(t) \eta_A(t) \right\rangle$ due to the onset of excited donor and acceptor fluctuations $\left\langle \xi^2_D(t) \right\rangle $ and $\left\langle \eta^2_A(t) \right\rangle $.
Those terms could have been derived independently from Eq.\ref{CorrEq1}, doing an \textit{a priori} Gaussian ansatz closure hypothesis for the third moments $\left\langle \sigma_\alpha(t)\sigma_\beta(t)\bsigma_j(t)\right\rangle$ and $\left\langle \sigma_\alpha(t)\bsigma_k(t)\bsigma_j(t)\right\rangle$.
We note, however, that the large-$\Omega$ asymptotic expansion of van Kampen \cite{van1992stochastic} is more suited to derive this result, since it recovers as a consequence of the approximation that the fluctuations are Gaussian in the macroscopic limit, without supposing it initially, nor truncating arbitrarily the ME.
Finally, the last term of Eq.\ref{CME_Hom_3} is the one that dominates at short times, since 
$\frac{d}{dt}\left\langle \xi_D(t) \eta_A(t) \right\rangle \approx -\tilde{k} \phi^{\rm{(mf)}}_{A}(0) \phi^{\rm{(mf)}}_{D^*}(0)=-\tilde{k}$ when $t \rightarrow 0^+$.
\textit{This term basically describes an anti-correlation between the populations of excited donor and acceptor molecules.} 
This is due to the intrinsic mechanism of RET reaction, for which each time an excited donor molecule $D^*$ transfers its energy to an acceptor in the ground-state, the latter gets excited to $A^*$.
This anti-correlation is beyond the mean-field approximation which supposes statistical independence between the donor and acceptor populations. 
%

%
\section{Reaction kinetics of RET}
\label{FullKin}
%
%
\subsection{Regular square lattice}
\label{FullKinSquare}
In this section, we show the outcome of the exact MC calculation (see Sec.\ref{Kin}).
We compute the time-dependent average population of excited donors $\left\langle N_{D^*}(t)\right\rangle$ and acceptors $\left\langle N_{A^*}(t)\right\rangle$, during the RET reaction.
Comparison is made to the analytical results provided by the macroscopic mean-field equations (see Sec.\ref{Macro}) which are written $\left\langle N_{D^*} \right\rangle^{\rm{(mf)}}(t)$ and $\left\langle N_{A^*} \right\rangle^{\rm{(mf)}}(t)$.
The donor and acceptor molecules are located on the sites of the regular bi-dimensional square lattice of lattice parameter $2a$ (intermolecular distance $a$) shown in Fig.~\ref{fig:Fig1}-Left. 
%

%
%
%
\subsubsection{Limit $a>R_0$}
\label{Larger}
\begin{figure}[tbh]
\includegraphics[width=1.0\linewidth, angle=-0]{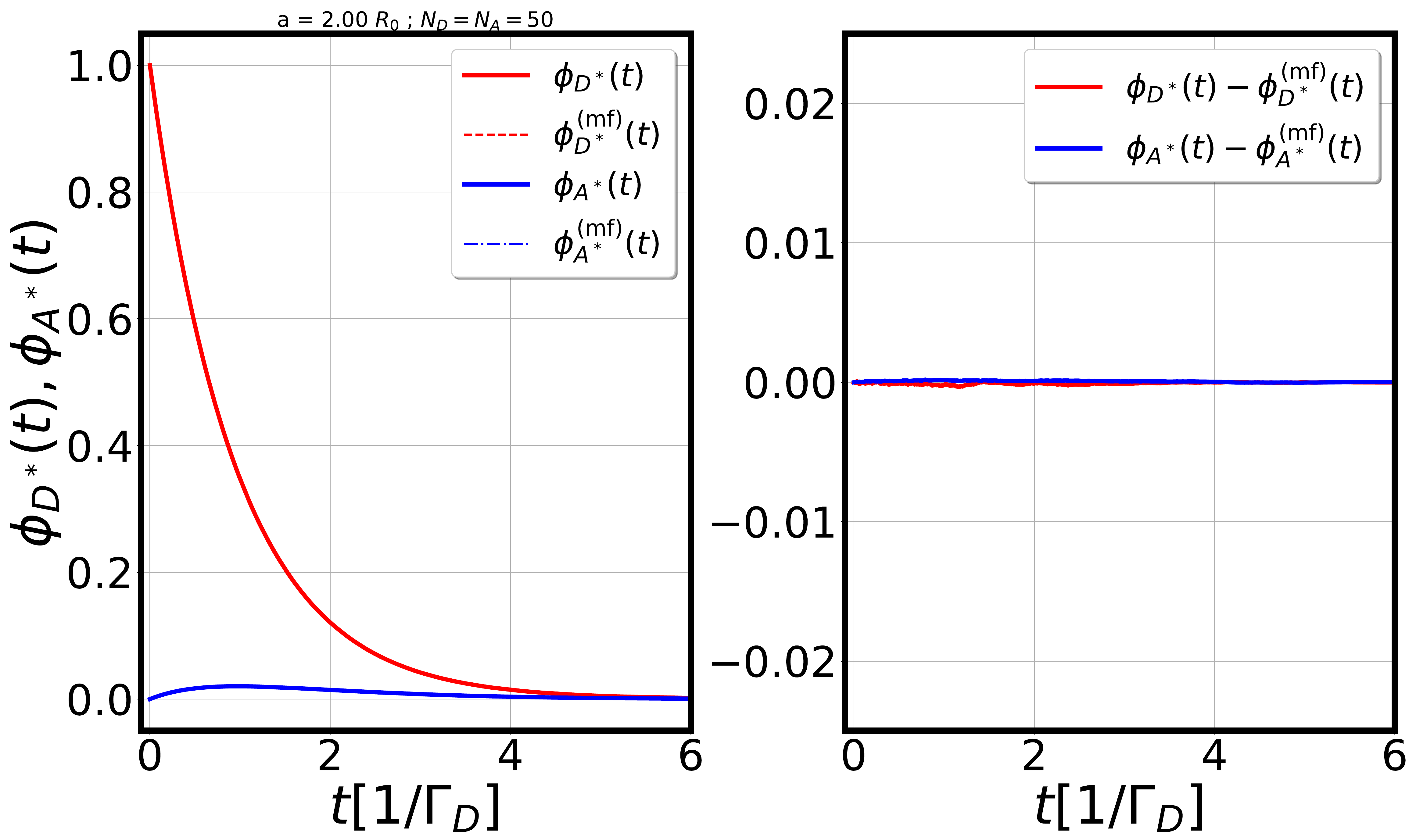}
\caption{
Left: Time-dependent populations of excited donors $\phi_{D^*}(t)=\left\langle N_{D^*}(t)\right\rangle/N_D$ (plain red curve) and acceptors $\phi_{A^*}(t)=\left\langle N_{A^*}(t)\right\rangle/N_A$ (plain blue curve) obtained from the MC calculation, in the case of a regular square lattice.
Comparison is shown with the macroscopic mean-field analytical results $\phi_{D^*}^{(\rm{mf})}(t)$ (dashed red curve) and $\phi_{A^*}^{(\rm{mf})}(t)$ (dash-dotted blue curve).
Right : Time-dependence of the contribution due to fluctuations $\phi_{D^*}(t)-\phi_{D^*}^{(\rm{mf})}(t)$ (red curve) and $\phi_{A^*}(t)-\phi_{A^*}^{(\rm{mf})}(t)$ (blue curve).
Parameters: $a=2.0 R_0$, $\Gamma_D=\Gamma_A\equiv \Gamma$ and $N_D=N_A=50$.
\textcolor{black}{The MC runs are averaged on $N_{traj}=10^5$ stochastic trajectories.}
}
\label{fig:Fig4}
\end{figure}
%
%
We present in the left panel of Fig.~\ref{fig:Fig4}, the outcome of the MC calculation for $\phi_{D^*}(t)\equiv\left\langle N_{D^*}(t)\right\rangle/N_D$ (plain red curve) and $\phi_{A^*}(t)\equiv\left\langle N_{A^*}(t)\right\rangle/N_A$ (plain blue curve), in the case $a=2R_0$.
In this regime, the distance $a$ between two nearest-neighbour molecules is larger than the F\"orster radius $R_0$.
\textit{Thus spontaneous emission is much faster than RET ($\Gamma \gg k $), and almost no energy-transfer occurs from the excited donors $D^*$ to the ground-state acceptors $A$.}
The average curve $\phi_{D^*}(t)$ for the donors, thus decays exponentially in time due to fluorescence and the corresponding curve $\phi_{A^*}(t)$ for the acceptors is almost zero, due to the fact that acceptor molecules are almost never excited.
%
%
For comparison, we show on the same plot, the output of the analytical macroscopic mean-field results $\phi_{D^*}^{\rm{(mf)}}(t)$ (dashed red curve) and $\phi_{A^*}^{\rm{(mf)}}(t)$ (dash-dotted blue curve), obtained from Eqs.\ref{Macro8},\ref{Macro9}, with $\tilde{k}$ computed using Eq.\ref{Macro11}. 
\textit{The analytical curves have thus no fitted parameter, and \textcolor{black}{are in excellent agreement} with the outcome of the MC calculations.}
%
%

%
We show in the right panel of Fig.~\ref{fig:Fig4}, deviations of the mean-field approximation from the MC calculation, due to fluctuations and contained in the curves $\phi_{D^*}(t)-\phi^{\rm{(mf)}}_{D^*}(t)$ (red curve) and $\phi_{A^*}(t)-\phi^{\rm{(mf)}}_{A^*}(t)$ (blue curve). 
In both cases, those deviations are very weak. 
This is due to the fact that in this regime of vanishing RET, the mean-field solution becomes exact for describing relaxation of the populations \textcolor{black}{dominated} by fluorescence.
%

%
%
%
\subsubsection{Crossover region $a=R_0$}
\label{Cross}
\begin{figure}[tbh]
\includegraphics[width=1.0\linewidth, angle=-0]{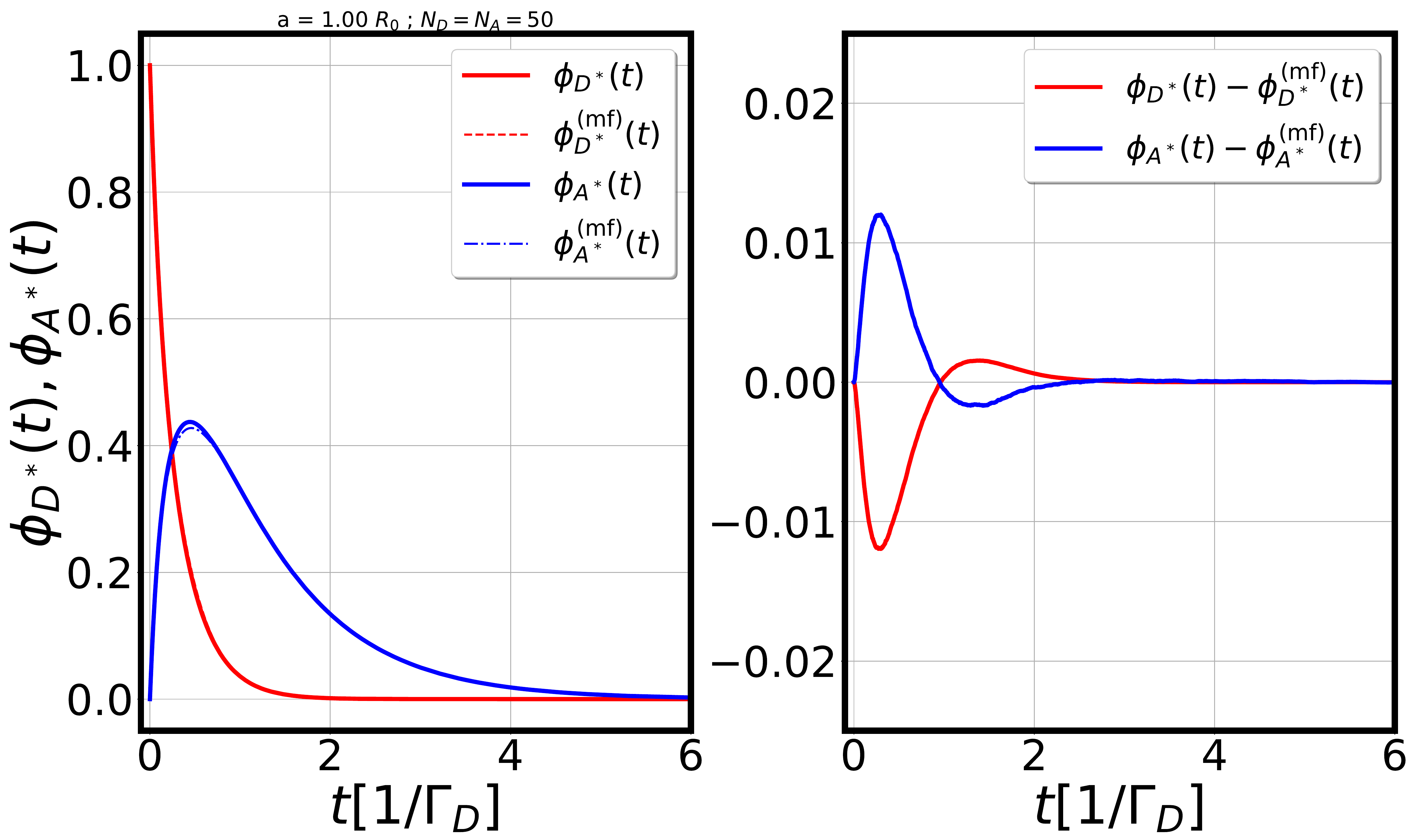}
\caption{
Same plot as in Fig.~\ref{fig:Fig4}, but with $a=R_0$.
}
\label{fig:Fig5}
\end{figure}
We show in Fig.~\ref{fig:Fig5}(left-panel), the same plots as in Fig.~\ref{fig:Fig4},   
but in the crossover regime where the distance $a$ between two neighbour $D$ and $A$ molecules equates the F\"orster radius $R_0$.
\textit{In this regime, the rate of spontaneous emission is of the same order of magnitude as the rate of RET ($\Gamma \equiv k $), so that there is competition between both processes and a significant energy-transfer occurs towards the acceptors.} 
In contrast to Fig.~\ref{fig:Fig4}, we see now that, while the curve of $\phi_{D^*}(t)$ (plain red) is still a monotonic decreasing function of time, the curve of $\phi_{A^*}(t)$ (plain blue) is no more zero and reaches a maximum at time $t \approx 1/\Gamma$. 
We understand this evolution of $\phi_{A^*}(t)$ with the following simple argument: at short-times 
($t \leq 1/\Gamma$), the donor molecules $D^*$ which are all initially excited, start to decay by fluorescence and a bit later to undergo energy-transfer processes, thus exciting the neighbouring acceptor molecules and increasing their relative number $\phi_{A^*}(t)$.
Once almost all $D^*$ molecules have decayed to their ground-state at times $t > 1/\Gamma$, the remaining population of excited acceptors $A^*$ (generated by RET) will relax back to their ground-state (no more RET is then possible), thus explaining the decrease at long-times of $\phi_{A^*}(t)$ towards zero.
The deviations of the mean-field curves with respect to the exact MC calculation are shown in the right panel of Fig.~\ref{fig:Fig5}.
The mean-field approximation is still a very good one, but contrary to Fig.~\ref{fig:Fig4}, the deviations due to fluctuations are no more completely negligible. 
In particular, at short times $t \ll 1/\tilde{k}$, those deviations become weakly negative, namely $\phi_{D^*}(t)-\phi_{D^*}^{\rm{(mf)}}(t) < 0$ (red curve).
Indeed the initial slope of this curve is proportional to the initial fluctuations $\frac{d}{dt}\left\langle \xi_D(t) \eta_A(t) \right\rangle \approx -\tilde{k}$, given by Eq.\ref{CME_Hom_3}.
\textit{The latter exhibit a sizable anti-correlation effect due to the anti-correlated mechanism of disappearance of $D^*$ \textcolor{black}{and appearance of} $A^*$ molecules in the elementary RET process.}
At longer-times $t > 1/\tilde{k}$, the onset of excited donor and acceptor fluctuations $\left\langle \xi^2_D(t) \right\rangle $ and $\left\langle \eta^2_A(t) \right\rangle $ in Eq.\ref{CME_Hom_3} start to couple to the curve $\phi_{D^*}(t)-\phi_{D^*}^{\rm{(mf)}}(t)$ and leads to its increase (with even a small overshoot above zero). 
At long-times $t \gg 1/\Gamma$, the curve $\phi_{D^*}(t)-\phi_{D^*}^{\rm{(mf)}}(t)$ goes back to zero due to the regression of fluctuations in Eq.\ref{CME_Hom_3}. 
In summary, the mean-field curves $\phi_{D^*}^{\rm{(mf)}}(t)$ and $\phi_{A^*}^{\rm{(mf)}}(t)$ constitute very good approximations to the MC calculation in the large-$N$ limit, which thus validates \textit{a posteriori} the use of this approximation.
Deviations to it are due to \textcolor{black}{mesoscopic} fluctuations that incorporate information about the microscopic RET process and related "three-body" competition mechanism between donor and acceptor molecules.
%

%
%
%
\subsubsection{Limit $a<R_0$}
\label{Small}
\begin{figure}[tbh]
\includegraphics[width=1.0\linewidth, angle=-0]{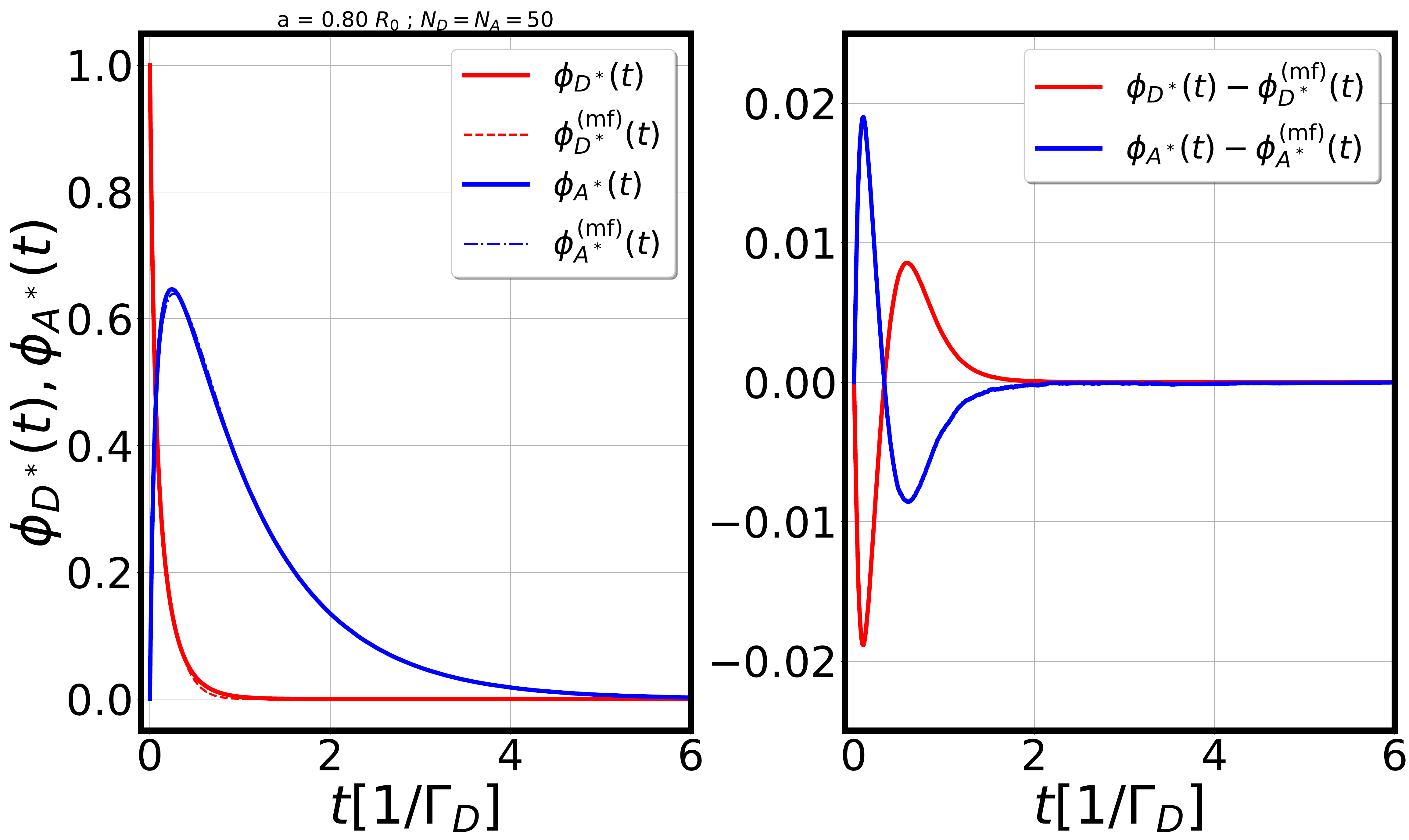}
\caption{
Same plot as in Fig.~\ref{fig:Fig4}, but with $a=0.8 R_0$.
}
\label{fig:Fig6}
\end{figure}
Finally, we show in Fig.~\ref{fig:Fig6} left panel, the same plots as in Fig.~\ref{fig:Fig4}, but in the opposite regime where the distance $a$ between two neighbour $D$ and $A$ molecules is lower than the F\"orster radius $R_0$.
\textit{In this regime, spontaneous emission happens with a much lower rate than RET ($\Gamma \ll k $), so that spontaneous emission from the excited donors is quenched at short times, when the donors decay faster through energy-transfer towards the acceptors.} 
At short-times ($t \leq 1/\tilde{k}$), $\phi_{A^*}(t)$ (plain blue curve) thus increases fast to a large value while $\phi_{D^*}(t)$ (plain red curve) decreases as fast to zero.
At longer times ($t \geq 1/\tilde{k}$), the excited acceptors decay back to ground-state by spontaneous emission on a time-scale given by $1/\Gamma$.
We show in the right panel of Fig.~\ref{fig:Fig6}, the related deviations of the mean-field approximation from the exact MC calculation.
The red curve $\phi_{D^*}(t)-\phi_{D^*}^{\rm{(mf)}}(t)$ shows a similar time-dependence as in Fig.~\ref{fig:Fig5}, but with faster evolution and higher deviations in amplitude. 
This is due to the value of $\tilde{k}$ that is larger in this regime than in Fig.~\ref{fig:Fig5}, thus amplifying effects of \textcolor{black}{mesoscopic} fluctuations due to the RET process.
%

%
%
\subsection{Disordered square lattice}
\label{FullKinRand}
\begin{figure}[tbh]
\includegraphics[width=1.0\linewidth, angle=-0]{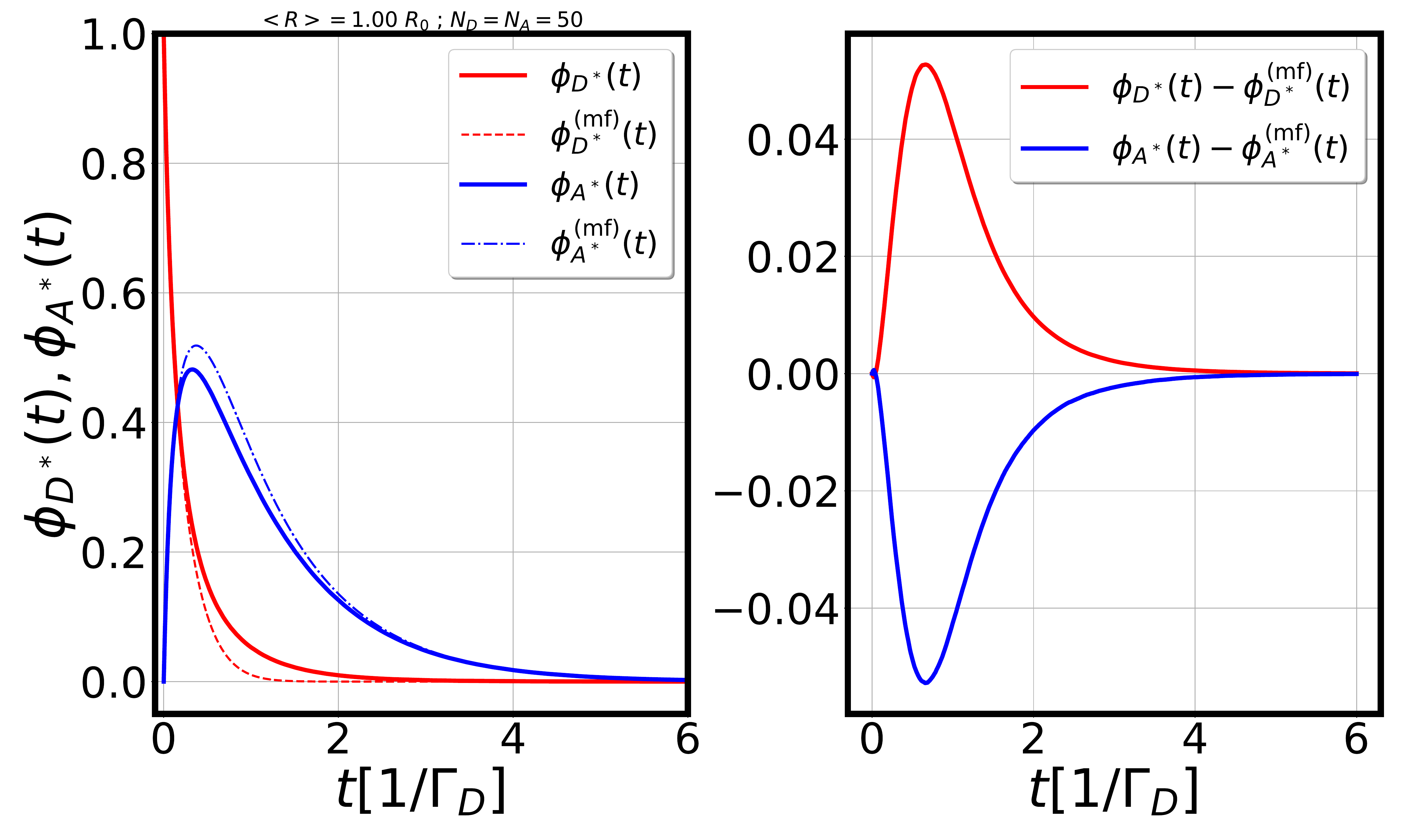}
\caption{
Left: Time-dependent populations of excited donors $\phi_{D^*}(t)=\left\langle N_{D^*}(t)\right\rangle/N_D$ (plain red curve) and acceptors $\phi_{A^*}(t)=\left\langle N_{A^*}(t)\right\rangle/N_A$ (plain blue curve) obtained from the MC calculation, in the case of a random square lattice.
Comparison is shown with the macroscopic mean-field analytical results $\phi_{D^*}^{(\rm{mf})}(t)$ (dashed red curve) and $\phi_{A^*}^{(\rm{mf})}(t)$ (dash-dotted blue curve).
Right : Time-dependence of the contribution due to \textcolor{black}{fluctuations} of $\phi_{D^*}(t)-\phi_{D^*}^{(\rm{mf})}(t)$ (red curve) and $\phi_{A^*}(t)-\phi_{A^*}^{(\rm{mf})}(t)$ (blue curve).
\textcolor{black}{Parameters: $\left\langle R \right\rangle=1.0 R_0$, $R_c=0.8R_0$, $\Gamma_D=\Gamma_A\equiv \Gamma$, $N_D=N_A=50$ and $N_{traj}=10^5$.}
}
\label{fig:Fig7}
\end{figure}
For completeness, we show in the left panel of Fig.~\ref{fig:Fig7} the curves of $\phi_{D^*}(t)$ and $\phi_{A^*}(t)$, in the case of a sample of molecules dispersed in a disordered square-lattice (as in Fig.~\ref{fig:Fig1}-Right), with $\left\langle R \right\rangle=1.0 R_0$ and $N_D=N_A=50$.
This range of parameters corresponds roughly to the same crossover regime as in Fig.~\ref{fig:Fig5} regarding the RET dynamics, and implies that both plots are qualitatively similar.
For the case of Fig.~\ref{fig:Fig7} however, the mean-field approximation (dashed and dash-dotted curves) shows larger deviations in absolute value with respect to the exact MC calculation (plain curves).
In the right panel of Fig.~\ref{fig:Fig7}, the initial negative contribution to $\phi_{D^*}(t)-\phi_{D^*}^{\rm{(mf)}}(t)$ is still there but very tiny compared to Fig.~\ref{fig:Fig5}, while the positive overshoot dominates the fluctuation signal with a larger amplitude. 
This lower accuracy of the mean-field approximation or increased role of fluctuations in the case of a disordered network, is due to the fact that one needs a larger number of molecules to explore several configurations of disorder and thus reach the macroscopic limit.
One way to recover a good accuracy of the mean-field results in this case would be to perform an average of the $\phi_{D^*}(t)$ and $\phi_{A^*}(t)$ curves over many different configurations of disorder (not shown here): this would mimic the self-averaging of disorder configurations upon increasing the sample size. 
%

\section{Conclusion}
\label{Conclusion}
In this paper, we have investigated in depth, the stochastic dynamics developing in two-dimensional samples containing donor and acceptor molecules, in presence of resonance energy transfer reactions and fluorescence. 
We have computed the populations of excited donor and acceptor molecules using a numerically exact kinetic Monte-Carlo approach and compared it to a non-linear mean-field approximation.
We derived exact kinetic equations describing the RET kinetics and predicted that, within mean-field approximation, the effective rate of RET $\tilde{k}$ in the macroscopic limit, depends in a non-trivial way of the molecular concentration and on the spatial distribution of pairs of molecules inside the sample. 
We showed that the rate $\tilde{k}$ scales with $\left\langle R \right\rangle^{-6}$ in the case where the molecules are put on the apex of an ordered square lattice, with $\left\langle R \right\rangle$ is the average distance between exited donors and their neighbor acceptors involved in the RET process. 
%
%
In contrast to this, we predicted that $\tilde{k}$ scales with $\left\langle R \right\rangle^{-2}$ in the case of a disordered sample, due to a different distribution of neighbouring acceptor molecules around each excited donor molecule.
Finally, we investigated the full time-dependence of the population kinetics and showed a good agreement with the mean-field approximation. 
Deviations of the latter compared to the exact MC calculations were shown to be due to fluctuations and correlations developing at the microscopic level and involving up to three molecules in the elementary RET process. 
Those deviations should be important in finite-size clusters or mesoscopic samples of molecules, as currently seen in biology. 
We note that our predictions for the scaling of the effective RET dynamics with molecular concentration should be observable in state of the art transient-fluorescence or pump-probe experiments. 
In the former case, additional effects at ultra-short times (shorter that the exciton-exciton annihilation time) should be observable that result from the interplay between the time-dependent drive by the pump followed by the slower relaxation dynamics due to fluorescence, exciton-exciton relaxation, and RET. 
The theoretical description of such effects in spatially inhomogeneous and anisotropic samples (position and orientational disorder of the molecular dipoles) is still an open issue, that would necessitate a generalization of our approach towards the use of a generalized master equation \cite{PhysRevB.9.5279}.
We hope that our results will be of interest to drive future theoretical and experimental investigations of RET in nanoscale systems, the latter physical process still exhibiting surprising and rich effects in out-of-equilibrium situations.
%

%
\begin{acknowledgments}
R.A acknowledges fruitful discussions with Thomas Gu\'erin about the large-$\Omega$ expansion used in Sec.\ref{Fluctuations}.
R.A. acknowledges financial support by Agence Nationale de la Recherche project CERCa, ANR-18-CE30-0006 and IdEx of the University of Bordeaux/Grand Research Program GPR LIGHT.
\end{acknowledgments}
%

%
%
\bibliography{biblio}
%

%
%
\appendix
%
%
%
\section{Uncertainty in the MC calculations}
\label{Unc_MC}
\begin{figure}[tbh]
\includegraphics[width=1.0\linewidth, angle=-0]{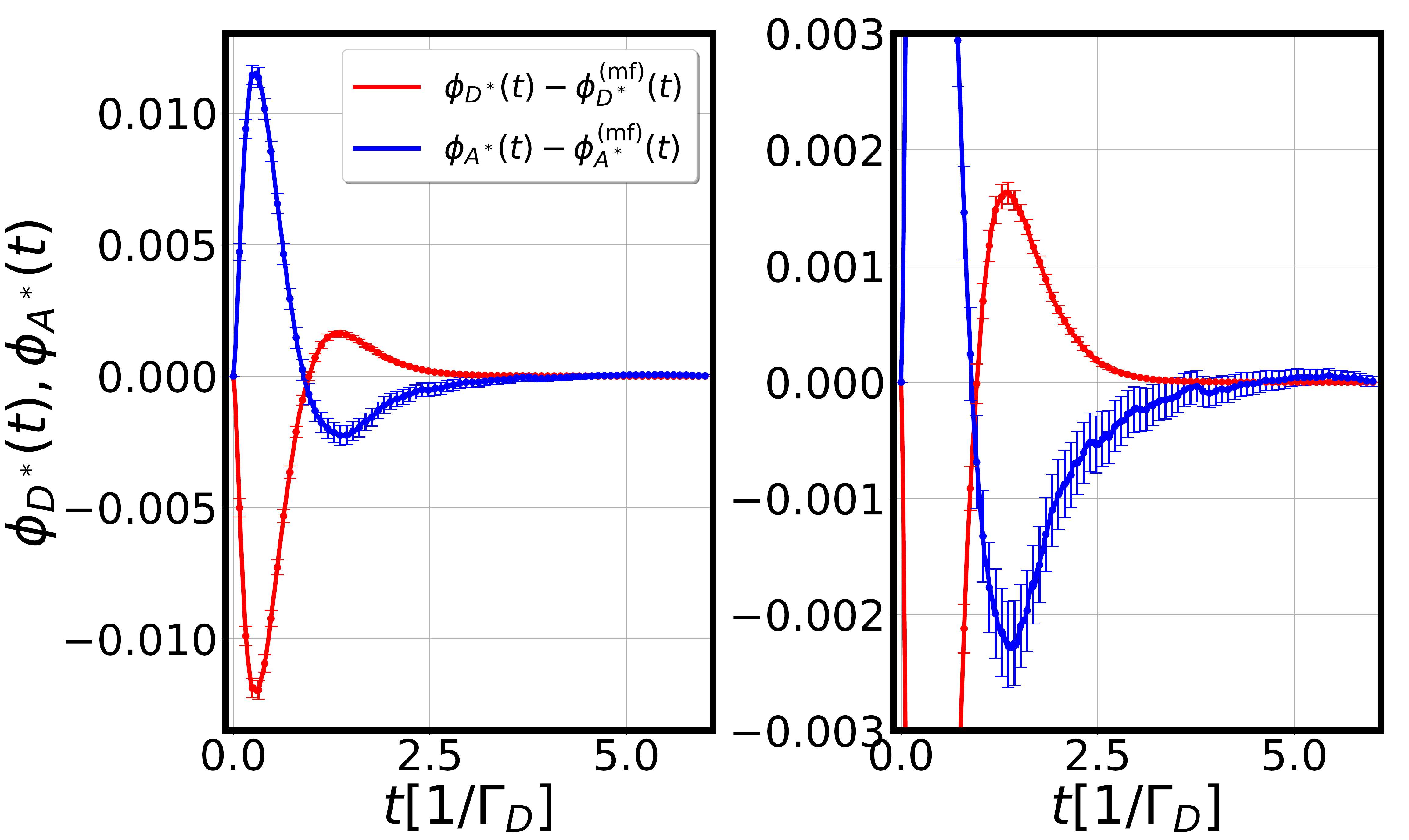}
\caption{
Left : Time-dependence of $\phi_{D^*}(t)-\phi_{D^*}^{(\rm{mf})}(t)$ (red curve) and $\phi_{A^*}(t)-\phi_{A^*}^{(\rm{mf})}(t)$ (blue curve),
for the same parameters as in Fig.~\ref{fig:Fig5}.
The MC runs are averaged on $N_{traj}=10^5$ time-trajectories.
We estimate here the statistical error in the MC runs, with error bars representing
two standard deviations. 
}
\label{fig:Fig8}
\end{figure}
In this Appendix, we estimate the statistical uncertainty in our MC calculations. 
Considering for instance the time-dependent average value in the population of excited donor molecules $\left\langle N_{D^*}(t) \right\rangle$, the finite number $N_{traj}$ of stochastic trajectories taken into account in performing the computation of $\left\langle N_{D^*}(t) \right\rangle$, induces a statistical uncertainty that we estimate to be of order $\Delta N_{D^*}(t)/\left\langle N_{D^*}(t) \right\rangle \approx 1/\sqrt{N_{traj}}$, with $\Delta^2 N_{D^*}(t) = \left\langle \left\lbrack N_{D^*}(t) - \left\langle N_{D^*}(t) \right\rangle \right\rbrack^2 \right\rangle$ the variance in the population of excited donor molecules (computed with the MC calculation).
The chosen value of $N_{traj}=10^5$ enables to reach a statistical uncertainty of the MC calculation of order $\Delta N_{D^*}(t)/\left\langle N_{D^*}(t) \right\rangle \approx 0.3\%$.
We present in Fig.~\ref{fig:Fig8}-Left (and Fig.~\ref{fig:Fig8}-Right for a zoom) the computed error-bars representing two standard deviations, in the case of the observables $\phi_{D^*}(t)-\phi_{D^*}^{(\rm{mf})}(t)$ (red curve) and $\phi_{A^*}(t)-\phi_{A^*}^{(\rm{mf})}(t)$ (blue curve), with the same parameters as in Fig.~\ref{fig:Fig5} of Sec.\ref{Cross}.
The error-bars are sufficiently low to resolve correctly the observables $\phi_{A^*}(t)-\phi_{A^*}^{(\rm{mf})}(t)$ and $\phi_{D^*}(t)-\phi_{D^*}^{(\rm{mf})}(t)$.
%

%
%
\section{Uncertainty in computing $\tilde{k}^{(\rm{rand})}$}
\label{Unc_k_mf}
\begin{figure}[tbh]
\includegraphics[width=1.0\linewidth, angle=-0]{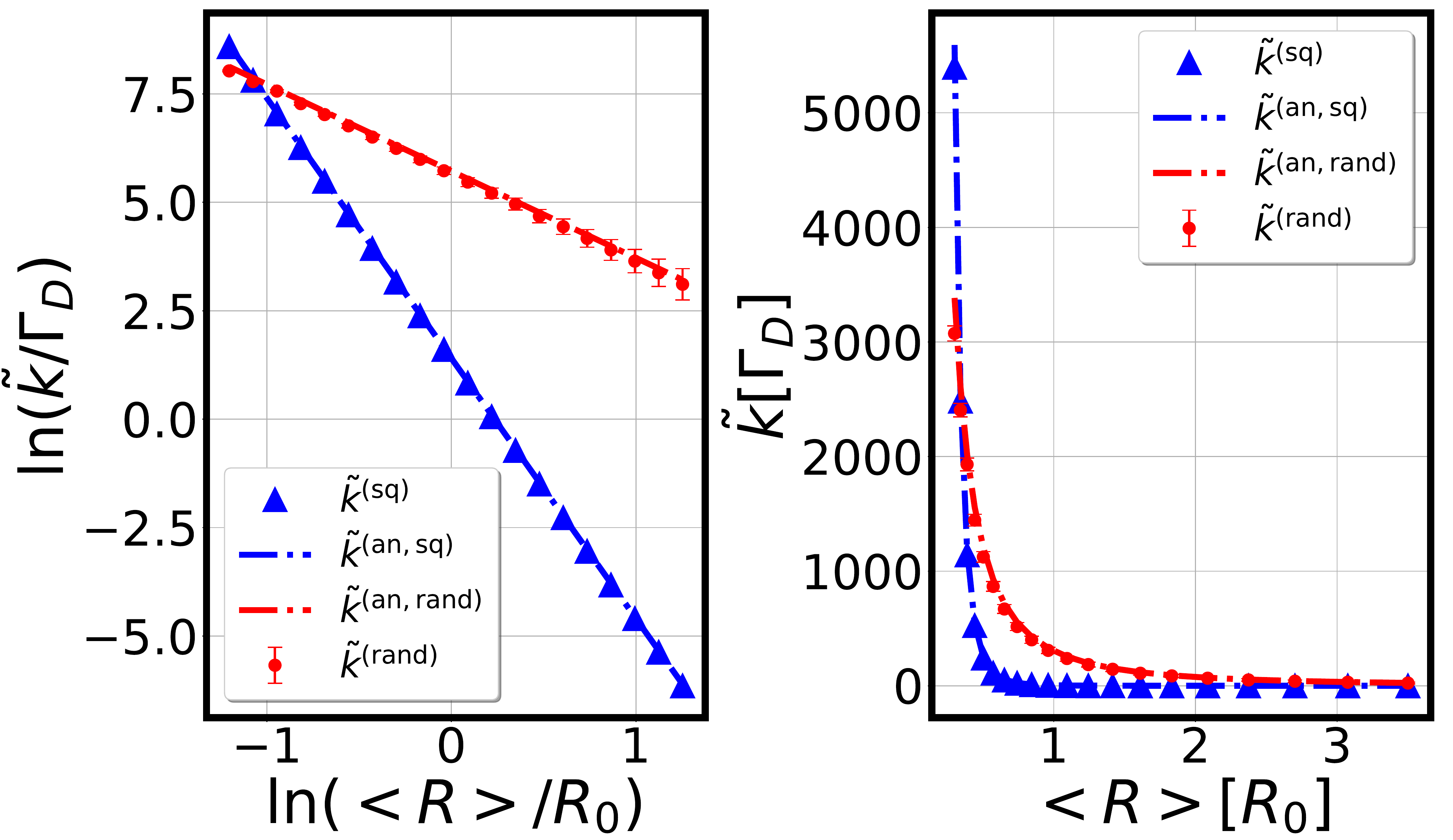}
\caption{
Same as Fig.~\ref{fig:Fig3}, but with a larger number of molecules $N_D=N_A=N/2=450$, and a shorter cutoff-length $R_c=0.3 R_0$
%
to compute $\tilde{k}^{(\rm{rand})}$. 
}
\label{fig:Fig9}
\end{figure}
We complement in this Appendix the discussion about the uncertainty in the numerical evaluation of the mean-field rate of RET $\tilde{k}^{(\rm{rand})}$ performed in Sec.\ref{FRETmf}.
We show in Fig.~\ref{fig:Fig9}, the same calculation of $\tilde{k}^{(\rm{rand})}$ performed in Fig.~\ref{fig:Fig3}, but with a larger number of molecules $N_D=N_A=N/2=450$, and a shorter cutoff-length $R_c=0.3 R_0$.
%
%
The red error bars are lower than the ones of Fig.~\ref{fig:Fig3}, essentially due to a larger number of molecules inside the sample, the former thus being closer to the large-$N$ limit for which the mean-field scaling law in Eq.\ref{PerfDis2} is expected to hold.  
The scaling law found for $\tilde{k}^{(\rm{rand})}$ as a function of $<R>$ is found to be the same as the one found in Sec.\ref{FRETmf}, thus strengthening our conclusion.
\end{document}